\documentclass[12pt]{article}

\usepackage{graphicx}
\usepackage{amssymb}
\usepackage{epstopdf}
\usepackage{authblk}
\usepackage{dsfont}

\usepackage{graphicx,epsfig}
\usepackage{graphics,dcolumn,bm,epic, eepic,fleqn,float}
\usepackage{amssymb,amsmath,amsfonts,multirow,rotate,color}
\usepackage{soul}
\usepackage[hyphens]{url}

\definecolor{Myorange}{cmyk}{0,0.42,1,0}

\title{An Algebraic Topological Method for Multimodal Brain Networks Comparisons}
\author[1,2,4]{Tiago Simas\thanks{ts526@cam.ac.uk}}
\author[3]{Mario Chavez}
\author[2]{Pablo Rodriguez}
\author[4]{Albert Diaz-Guilera}
\affil[1]{Department of Psychiatry, University of Cambridge, Cambridge, UK}
\affil[2]{Telefonica I+D, Barcelona, Spain}
\affil[3]{CNRS-UMR-7225, H\^opital Piti\'e Salp\^etri\`ere, Paris, France}
\affil[4]{Departament de Fisica Fonamental, Universitat de Barcelona, Barcelona, Spain}

\begin{document}
\maketitle
\begin{abstract}
Understanding brain connectivity has become one of the most important issues in neuroscience. But connectivity data can reflect either the functional relationships of the brain activities or the anatomical properties between brain areas. Although one should expect a clear relationship between both representations it is not straightforward. Here we present a formalism that allows for the comparison of structural (DTI) and functional (fMRI) networks by embedding both in a common metric space. In this metric space one can then find for which regions the two networks are significantly different. Our methodology can be used not only to compare multimodal networks but also to extract statistically significant aggregated networks of a set of subjects. Actually, we use this procedure to aggregate a set of functional (fMRI) networks from different subjects in an aggregated network that is compared with the anatomical (DTI) connectivity. The comparison of the aggregated network reveals some features that are not observed when the comparison is done with the classical averaged network. 
\end{abstract}

\section{Introduction}
In the last decade, the use of advanced tools deriving from neuroimaging and complex networks theory have significantly improved our understanding of brain functioning \cite{BOOKOLAF}. Notably, connectivity-based methods have had a prominent role in characterising normal brain organisation as well as alterations due to various brain disorders~\cite{varela01, stam12, stam14}. Most of the recent works aim to quantify the role of connectivity in the communication abilities of neural systems. However, the very same notion of connectivity is controversial since data used in brain connectivity studies can reflect functional neural activities (electrical, magnetic or hemodynamic/metabolic) or anatomical properties~\cite{varela01, bullmore09}. Neuroanatomical connectivity is meant as the description of the physical connections (axonal projections) between two brain sites~\cite{bullmore09}, whereas functional connectivity is defined as the estimated temporal correlation between spatially distant neurophysiological activities such as electroencephalographic (EEG), magnetoencephalographic (MEG) functional magnetic resonance imaging (fMRI) or positron emission tomography (PET) recordings~\cite{varela01}. 

In recent years, the concept of ``brain networks'' is becoming fundamental in  neuroscience~\cite{stam07, bullmore09, stam12, stam14}. Within this framework, nodes stand for different brain regions (e.g. parcelated areas or recording sites) and links indicate the presence of an \emph{anatomical} path between those regions, or a \emph{functional} dependence between their activities. In the last years, this representation of the brain has allowed to visualize and describe its non-trivial topological properties in a compact and objective way. Nowadays, the use of network-based analysis in neuroscience has become essential to quantify brain dysfunctions in terms of aberrant reconfiguration of functional brain networks~\cite{stam07, stam12, stam14}. 

Experimental evidence has revealed, for instance, alterations in functional and anatomical brain networks in normal cognitive processes, across development, and in a wide range of neurological diseases (see~\cite{bullmore09, stam14} and references therein). Despite its evident interplay, comparison of anatomo-functional brain networks is not straightforward~\cite{deco11, nicosia14}. Theoretical studies provide support for the idea that structural networks determine some aspects of functional networks~\cite{deco11}, but it is less clear how the anatomical connectivity supports or facilitates the emergence of functional networks. Although nodes with similar connection patterns tend to exhibit similar functionality, the functionality of an individual neural node is strongly determined by the pattern of its interconnections with the rest of the network~\cite{nicosia14}.

Correspondences between functional and structural networks remains thus an active research area~\cite{honney07, honney09, honney10}. A better understanding of how anatomical scaffolds support functional communication of brain activities is necessary to better understand normal neural processes, as well as to improve identification and prediction of alterations in brain diseases.

In this paper we address this relationship between anatomical and functional connectivity. In previous studies, the correspondence of these networks has been often assessed by the difference in an Euclidean space of vectors containing connectivity measures such as the clustering coefficient, shortest path length, degree distribution, etc. Here, we propose a radically different framework for studying brain connectivity differences. Instead of extracting a vector of features for each network (anatomical or functional), we jointly embed all of them in a common metric space that allow straightforward comparisons. Before embed functional and anatomical networks into the common metric space, we aggregate group of subjects (e.g. functional networks) according to \cite{SCD} to obtain a group representation network. Therefore the method employed in this work allows to preserve in the aggregation connected components and to identify among different subjects, a common underlying network structure. 
Our approach may provide a useful insight for the analysis of multiple networks obtained from multiple brain modalities or groups (healthy volunteers \emph{versus} patients, for instance).

\section{Methods and Materials}
\subsection{fMRI and DTI data}
In this study we considered anatomical and functional brain connectivities 
(extracted from diffusion-weighted DW-MRI and fMRI data, respectively) defined on 
the same brain regions. Brain images were partitioned into the 90
anatomical regions ($N=90$ nodes of the networks) of the
Tzourio-Mazoyer brain atlas~\cite{TzourioMazoyer2002} using the automated anatomical labeling method.

The anatomical connectivity network is based on the connectivity matrix obtained by
Diffusion Magnetic Resonance Imaging (DW-MRI) data from 20 healthy
participants, as described in~\cite{DTIdata2008}. The elements of this
matrix represent the probabilities of connection between the 90 brain regions of interest. These
probabilities are proportional to the density of axonal fibers between
different areas, so each element of the matrix represents an
approximation of the connection strength between the corresponding
pair of brain regions.

The functional brain connectivity was extracted from BOLD fMRI resting
state recordings obtained as described in~\cite{valencia2009}. 
All acquired brain volumes were corrected for motion and differences in slice acquisition 
times using the SPM5\footnote{http://www.fil.ion.ucl.ac.uk} software package.
All fMRI data sets (segments of 5 minutes recorded from healthy
subjects) were co-registered to the anatomical data set and normalized
to the standard MNI (Montreal Neurological Institute) template image,
to allow comparisons between subjects. As for DW-MRI data, normalized
and corrected functional scans were sub-sampled to the anatomical
labeled template of the human brain~\cite{TzourioMazoyer2002}. Regional time series were estimated
for each individual by averaging the fMRI time series over all voxels
in each of the 90 regions. To eliminate low frequency noise (e.g.~slow scanner drifts) and higher frequency artifacts from cardiac and respiratory
oscillations, time-series were digitally filtered with a finite impulse response (FIR) filter with zero-phase distortion (bandwidth
$0.01-0.1$~Hz) as in~\cite{valencia2009}.

A functional link between two time
series $x_i(t)$ and $x_j(t)$ (normalized to zero mean and unit
variance) was defined by means of the linear cross-correlation
coefficient computed as $r_{ij} = \langle x_i(t)x_j(t)\rangle $, where
$\langle\cdot\rangle$ denotes the temporal average. For the sake of
simplicity, we only considered here correlations at zero lag. To
determine the probability that correlation values are significantly
higher than what is expected from independent time series, $r_{ij}(0)$
values (denoted $r_{ij}$) were firstly variance-stabilized by applying the Fisher's Z
transform
\begin{equation}
\label{nonNormalisedW}
Z_{ij} = 0.5\ln \left(\frac{1+r_{ij}}{1-r_{ij}} \right)
\end{equation}
Under the hypothesis of independence, $Z_{ij}$ has a normal
distribution with expected value 0 and variance $1/(df_{ij}-3)$, where $df$
is the effective number of degrees of freedom~\cite{Bartlett1946,
Bayley1946, Jenkins1968}. If the time series consist of independent
measurements, $df_{ij}$ simply equals the sample size, $N$.  Nevertheless,
autocorrelated time series do not meet the assumption of independence
required by the standard significance test, yielding a greater Type I
error~\cite{Bartlett1946, Bayley1946, Jenkins1968}. In presence of
auto-correlated time series $df$ must be corrected by the following
approximation $\frac{1}{df_{ij}}\approx \frac{1}{N} + \frac{2}{N}\sum_\tau \frac{N-\tau}{N}r_{ii}(\tau) r_{jj}(\tau)$, 
where $r_{xx}(\tau)$ is the autocorrelation of signal $x$ at lag $\tau$. 

\subsection{Networks Normalisation}
From Eq.~(\ref{nonNormalisedW}) our networks weights are in a non-normalised interval $Z_{ij}\in[a,b]\subset\mathds{R}$. In order to apply the framework described in \cite{SR}, we normalise our networks weights into the unit interval $I=[0,1]$ by means of a unique linear function:

\begin{equation}
\label{linearnorm}
w_{ij}=\frac{(1-2\epsilon) Z_{ij}+(2\epsilon-1)\cdot MIN(Z_{ij})}{MAX(Z_{ij})-MIN(Z_{ij})}+\epsilon
\end{equation}

\noindent where $\epsilon$ in general is set to $0.01$ in order to avoid merging and isolate vertices with weights at the boundaries of $Z_{ij}\in[a,b]$. As proved in \cite{SR}, since the normalisation is done by a unique linear function this does not affect networks properties.

\subsection{fMRI Networks aggregation and embedding}
Among many ways to aggregate a group of networks here we employed a topological algebraic way to aggregate a group of networks. The networks group possess the same nodes but different edges values and can mathematical be represented by a weighted graph $G=(N,E)$. $N$ is the set of nodes representing the brain ROI's ($N= 90$ in this study) and $E$ is the set of edges values (connections) between ROI's, e.g. $\forall e_{i,j}\in E: e_{i,j} \in [0,1]$ in the proximity space or $\forall d_{i,j}\in E: d_{i,j} \in [0,+\infty]$ in the distance space.

For mathematical notation simplicity, we denote a network with the same notation we use to the set of nodes $N$.  That is, a set of $n$ networks (e.g. group of subjects) is represented by $N_{k}$ with $k\in \{1,2,3.\dots, n\}$.

One possible way to aggregate a group of $n$ networks is simply by averaging the homologous edges values. Obtaining in this way a group representative network, $N^{*}$.

\begin{equation}
\label{average}
N_{i,j}^{*}=e_{i,j}^{*}=\frac{\sum_{k=1}^{n} e_{i,j}^{[k]}}{n}
\end{equation} 

\noindent where $e_{i,j}^{[k]}$ is the edge $e_{i,j}$ from network $N_k$.

\begin{figure}[h!]
\centering
\resizebox{0.95\textwidth}{!}{
\includegraphics{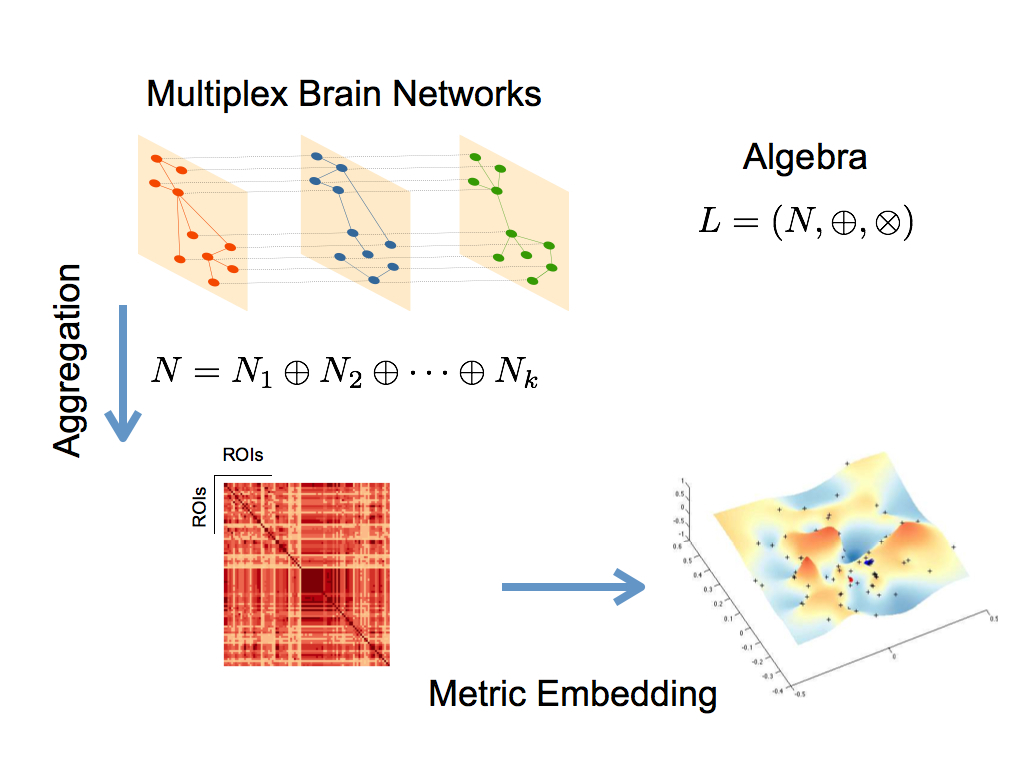}}
\caption{Schematic representation of the main steps for the described networks aggregation and metric embedding (defined here for the algebra $L$).\label{fig1}}
\end{figure}

Another way to aggregate networks as explained in Simas et. al. \cite{SCD}, is by considering all networks as a \emph{multilayer} network (often called \emph{multiplex}), which can be represented as fourth-order tensor \cite{SCD}. This tensor can be represented as a extended matrix \cite{SOLE}. The work of Simas and Rocha \cite{SR}, introduces a framework to aggregate networks in an algebraic way, relating it with fuzzy logic reasoning, and in \cite{SCD} this work was extended for multilayer networks. In order to work algebraically with networks we have to set an algebra (defined as a vector space equipped with a bilinear product). This algebra allows us to perform algebraic operations with networks in the same way we perform algebraic operations in other contexts with other algebras (such as adding and multiplying real numbers). In short a network can be represented by an adjacency matrix and a multilayer network by a tensor. Considering a set of tensors working under the algebra $L=(I,\oplus,\otimes)$, where the weights (tensor entries) of the tensors in $I\subseteq \bar{\mathds{R}}$ (subset of \emph{extended real line}) and $\oplus$ and $\otimes$ two binary operators we may represent a multilayer network with tensor $T$ in this algebra. In Simas et al. \cite{SCD} we have seen for the particular case of multiplex networks where layers are connected with weights $w_{i,i,L_k,L_j}=1$ (in the proximity space), the representative group network (e.g. functional) can be represented by $N^{*}$ in the distance space (see below and Eq. \ref{isomorphism}), as: 

\begin{equation}
\label{algebra}
N^{*} \equiv N_{1}\oplus N_{2} \oplus \dots \oplus N_{k}
\end{equation} 

\noindent and the respective embedding by the following equation:

\begin{equation}
\label{algebra2}
N_{embedded}=N^{*}\oplus N^{*2} \oplus \dots \oplus N^{*r}
\end{equation} 

\noindent where $N^{*}$ is defined in Eq.~(\ref{algebra}) and $r$, is the convergence parameter~\cite{SR,SCD}. Figure \ref{fig1} summarises the metric embedding of a multiplex network described above.

Embedding a network of networks or, in our specific case, a a multiplex fMRI network, allows us to determine which edges in the several layers contribute to the aggregation. We can therefore determine the subjects that contribute more/less or none to the aggregated network, and identify in each subject the sub-graphs for which they contribute more.

For our particular case, we embed our networks using the Metric Closure \cite{SR} defined by the algebra $L=([0,+\infty],min,+)$, where $\oplus=min$ and $\otimes=+$. The metric closure or metric embedding of a given network into a metric space, is a generalisation of All Pairs Shortest Paths Problem (APSP) as shown in Ref.~\cite{SR}, e.g. Johnson Algorithm can be used to calculate the metric closure~\cite{JOHNSON}.

Note that to calculate the metric closure (or Johnson algorithm) of a network we have to translate our networks from a proximity space into a distance space. There are many possible mappings to map a similarity space into a distance space, see \cite{SR}. Applying Eq.~(\ref{isomorphism}) to all network weights, $w_{i,j}\in[0,1]$ (for more details see \cite{SR}), we obtain the isomorphic distance network with weights $d_{i,j}\in[0,+\infty]$.

\begin{equation}
\label{isomorphism}
d_{i,j}=\frac{1}{w_{i,j}}-1
\end{equation} 

Embedding networks or multilayer networks allows us: (a) to detected clusters of nodes in a high-dimensional topological spaces, and by projecting the algebraic high-dimensional embedding into 3D, it allows to perform exploratory networks analysis and; (b) to preserve the multilayer sub-structures across layers/subjects, better than other aggregations methods, as we show for the specific case of "simple" averaging (equation \ref{average}).  

Next we compared both methods of aggregation, "simple" averaging (eq. \ref{average}) and algebraic aggregation (eq. \ref{algebra2} according to \cite{SCD}) of our fMRI networks respectively, using our proposed method of embedding and comparison networks.

\subsection{Multimodal Networks comparison}
In general networks have been compared using statistical measures of local and global properties of networks, such as: clustering coefficient, small-worldness, degree distributions, etc. We can find in the literature some examples of such techniques to compare multiple networks~\cite{bullmore09, stam14}. Our approach in this work is different. After embedding networks into the same metric space defined by the applied algebra, in our case  $L=([0,+\infty],min,+)$, we are able to compared them topologically. However, since networks generally come from different modalities (e.g. fMRI and DTI) it requires a previous step. We need to normalise the embedded edge weights distributions from the different modalities to the same average and variance to remove scale factors.  One possible way to normalise both distributions, if we assume normality, is by calculating the z-score of the edge weights distributions (zero average and standard deviation set to the unit). 

The embedded networks represent a hyper-grid in a multi-dimensional space with dimension equal or below to the number of nodes. In order to simplify and have some visual insight we can downgrade linearly this multidimensional grid into a 3D grid. This can be achieved applying to the embedded networks any technique for dimensionality reduction such as linear/non-linear Multi-Dimensional Scaling (MDS). MDS procedures refer to a set of related ordination techniques used in information visualisation, in particular to display the information contained in a distance matrix~\cite{BOOKMDS}. These techniques guarantee, with a given distortion, that the relative distance between nodes is preserved in both multi-dimensional and low-dimensional reduction space. Plotting this low-dimensional grid (e.g. in 3D) we can use any statistical technique to fit a continuous surface into the data (see below Figs.~(\ref{fig3}, \ref{fig4})). Its is natural to think that the difference between two surfaces obtained from different networks will emphasise topologically differences between the two connectivities. In this work we performed this operation in the multi-dimensional space by subtracting homologous embedded edges weights and take the absolute value of both embedded hyper-grids. This is, we subtract homologous embedded edges pairwise according to the formula:

\begin{equation}
\label{hyper-grid}
\begin{aligned}
M= & |M_{fMRI^*}-M_{DTI}| \equiv & \\
     & \{e_{i,j}^{diff}: \forall e_{i,j}^{*} \in M_{fMRI^{*}} \land \forall e_{i,j}^{**} \in M_{DTI}:  e_{i,j}^{diff}=|e_{i,j}^{*}-e_{i,j}^{**} |\} &
\end{aligned}
\end{equation} 

\noindent $M$ is the difference grid in the multi-dimensional space. Because the $M$-grid represents the difference between the two grids from different modalities (see above), the relative distance between nodes in $M$ (given by Eq.~(\ref{hyper-grid})) should be concentrated at the origin if they are topological similar, otherwise widely distributed in the multi-dimensional space. Nodes at a distance from the origin of $k\times \sigma$ are statistical different of $k$ standard deviations. Moreover, since we z-scored both embedded edge distributions this give us some degree of statistical significance when we compare both networks. All nodes that relay outside of a hyper-sphere with centre at the origin with radius $R=k\times \sigma$, are statistical different. Here we had set $\sigma=1$ for both distributions (z-score variables are estimated from the distributions of the embedded weights). 

\begin{figure}[!ht]
\centering
\resizebox{0.95\textwidth}{!}{
\includegraphics{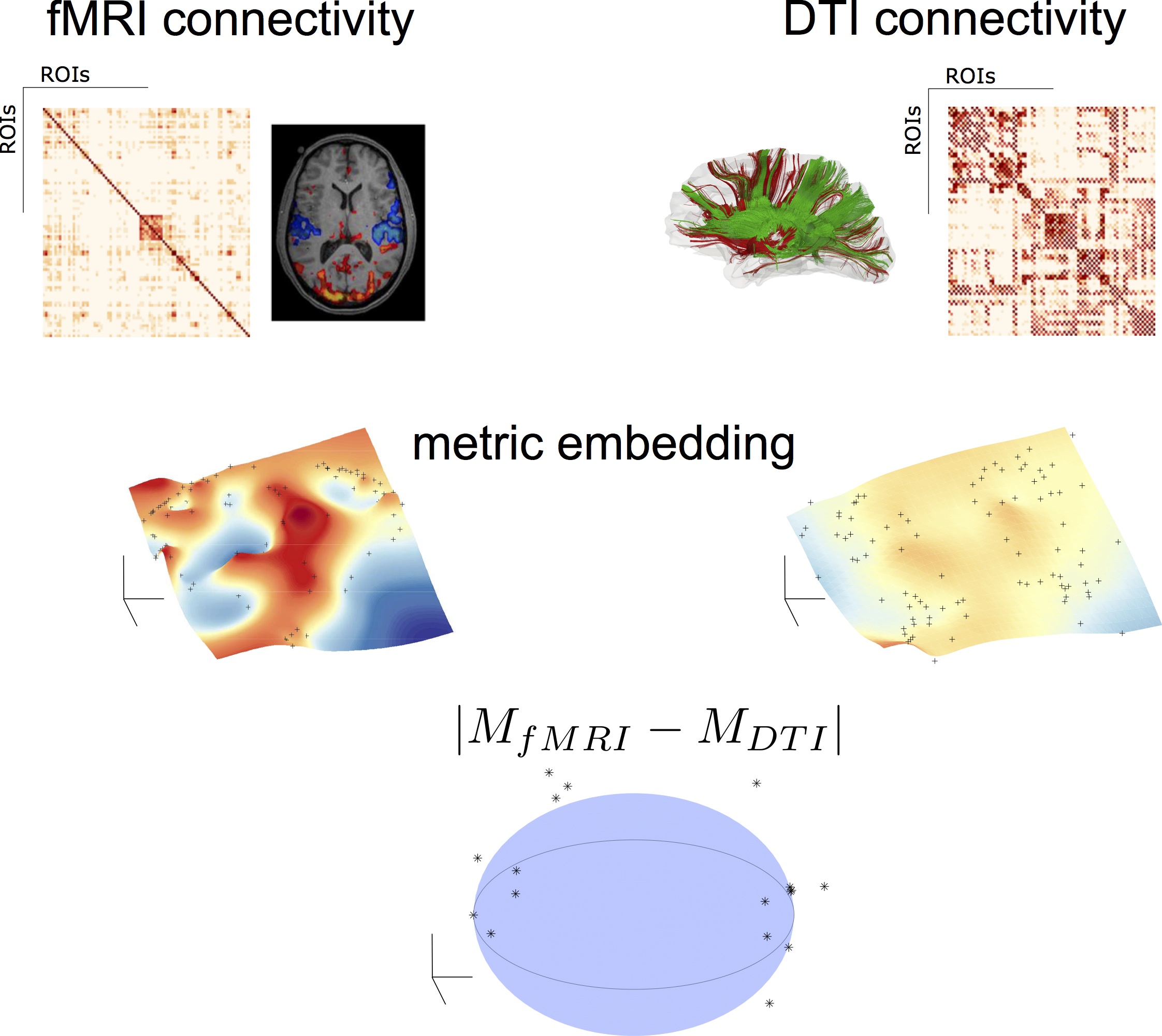}}
\caption{Topological algebraic networks comparison. Connectivity from different modalities (here fMRI and DTI) are firstly embedded (black dot points on the manifolds indicate the brain nodes) and then compared in a low-dimensional space. Black points outside the sphere correspond to nodes with a topological difference (at a given threshold) in the two modalities. \label{fig2}}
\end{figure}

Figure~\ref{fig2} illustrates this process. After applying to both fMRI$^{*}$ and DTI networks the same algebra and the metric embedding described above, both networks rely on the same metric space, therefore comparable. Topological differences can be visually seen in a linearly downgraded to 3D dimensions using a multi-dimensional scaling technique, which preserves the relative distance between points in the grid (nodes or brain areas). 

\section{Results}
\begin{figure}[h!]
\centering
\resizebox{0.95\textwidth}{!}{
\includegraphics{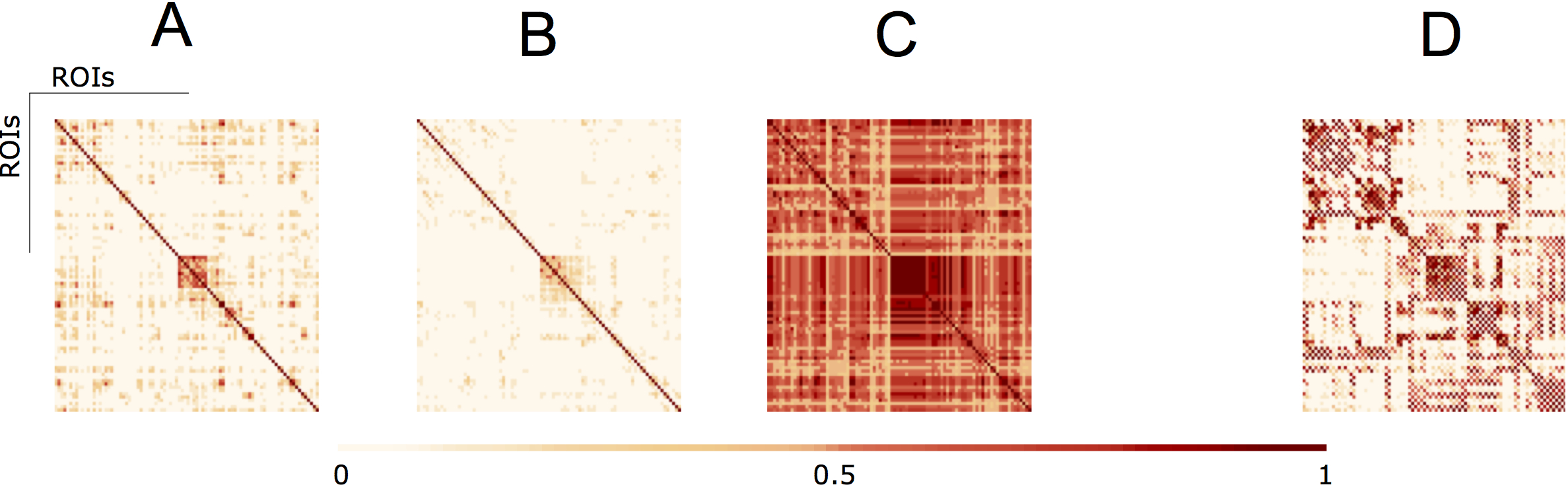}}
\caption{(A) fMRI single subject network (B) Average aggregated fMRI network (C) fMRI Algebraic Topologically aggregated (multiplex) network (D) DTI network.\label{fig3}}
\end{figure}

In figure~\ref{fig3} we illustrate the results of different aggregation procedures on the ensemble of fMRI networks. Compared with a fMRI connectivity matrix from a single subject (Fig.~\ref{fig3}(A)), one can notice the difference of a single averaging across subjects (Fig.~\ref{fig3}(B)) and our proposed algebraic topologically aggregated connectivity network (Fig.~\ref{fig3}(C)). It is clear that the averaging procedure tends to blur connectivity values between nodes. In contrast, the  topologically algebraic aggregation can preserve components that are common across subjects. As other multilinear algebra or tensor-based analysis, our approach provide a natural mathematical framework for studying connectivity data with multidimensional structure. For illustrative purposes, we also show the DTI connectivity matrix in figure~\ref{fig3}(D)). It worths noticing the similarity of the anatomical connectivity structure with the aggregated (multiplex) connectivity obtained in figure~\ref{fig3}(C). 
Moreover, since each layer encodes the functional network for a given subject, each subject contributes to the tensor aggregation/embedding with some or none connections (edges), as depicted in metric closure, Eq. (\ref{algebra2}). If a layer do not contribute for the aggregation/embedding, we may consider this layer (subject network) as an outlier. Moreover, we are also able to identify the specific sub-network contribution (edges) of a given layer to the aggregation/embedding. 

\begin{figure}[!ht]
\centering
\resizebox{0.9\textwidth}{!}{
\includegraphics{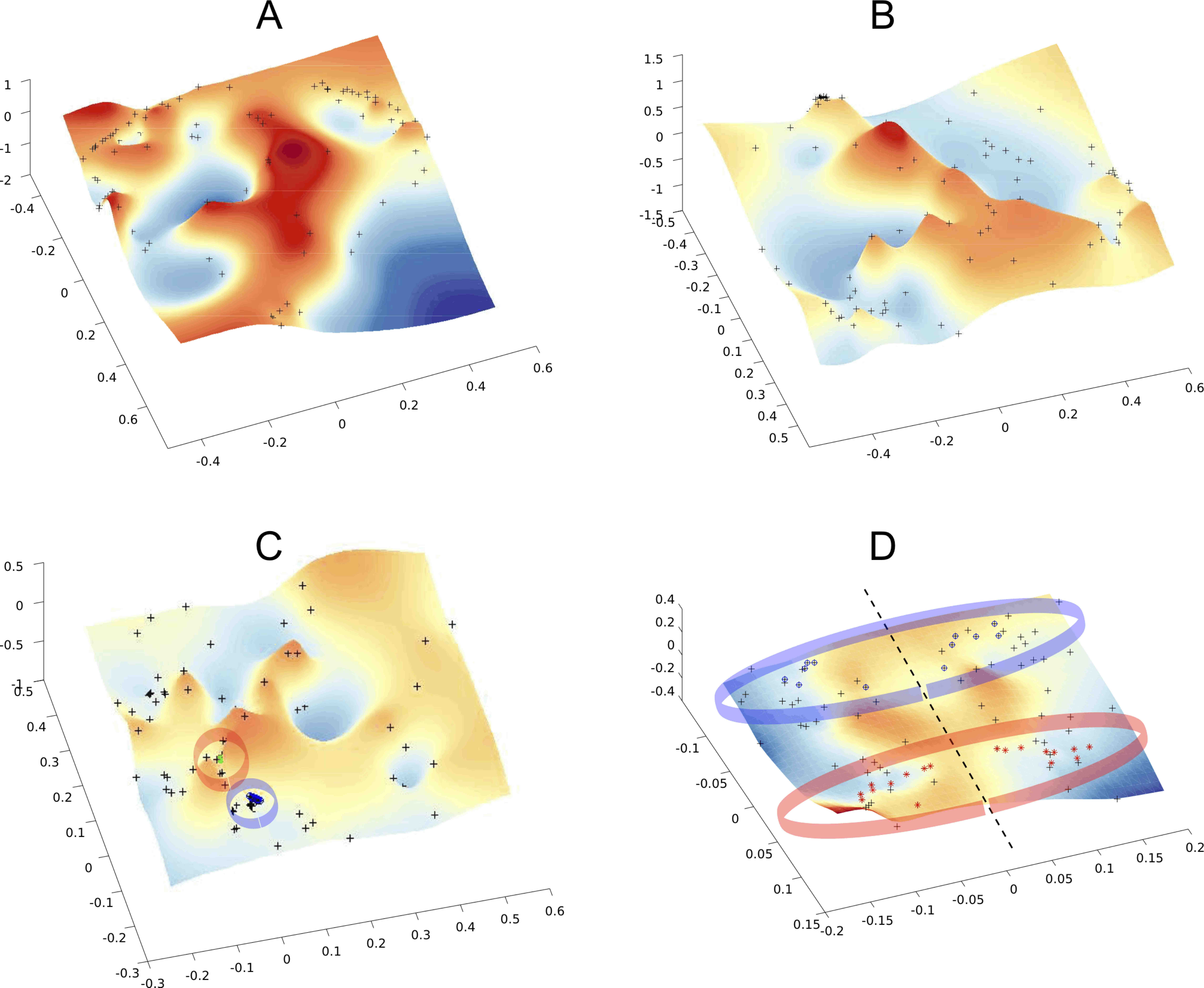}}
\caption{Multi-Dimentional Scaling (MDS) of the embedded networks (a) fMRI single subject  (b) fMRI average embedded network (c) fMRI Algebraic Topological aggregation (multiplex) embedded network (d) DTI embedded network. Black dots indicate the embedded nodes. In plots (C-D), blue and red points indicate the groups of brain areas discussed in the text. \label{fig4}}
\end{figure}

Low-dimensional embeddings of different aggregated networks are illustrated in figure~\ref{fig4}. High-dimensional data, such as the information contained in the distance matrix obtained for the different networks, can be difficult to interpret. Here, multidimensional scaling (MDS) was used for visualising the level of similarity of individual nodes of each -aggregated- network. The MDS algorithm aims to place each node in a low dimensional space such that the between-nodes distances are preserved as well as possible.  This representation into a low-dimensional space enables an exploratory analysis and makes data analysis algorithms more efficient.  Indeed, from the different plots of figure~\ref{fig4} one can identify brain areas that are topologically close in the aggregated network as those points that are close on the 3D grid. This is clearly illustrated by the MDS representation of the multiplex functional network (Figure~\ref{fig4} (C)). Nodes from the occipital regions form a compact group of nodes topologically close (with similar connectivity structure), as revealed by the blue points depicted on Figure~\ref{fig4} (C). We also notice that a compact group of nodes is formed by regions of the temporal lobe, putamen and insula, which are indicated by the red circle. Similarly, the anatomical network in Figure~\ref{fig4} (D) clearly displays a natural organisation, i.e. nodes of the two hemispheres lie on both sides of the dotted black line. Further,  nodes from occipital regions in the anatomical network, indicated by the blue circles (including calcarina, cuneus, precuneus, \ldots), are distantly located from the group of frontal brain areas indicated by the red marks.

\begin{figure}[!ht]
\centering
\resizebox{0.9\textwidth}{!}{
\includegraphics{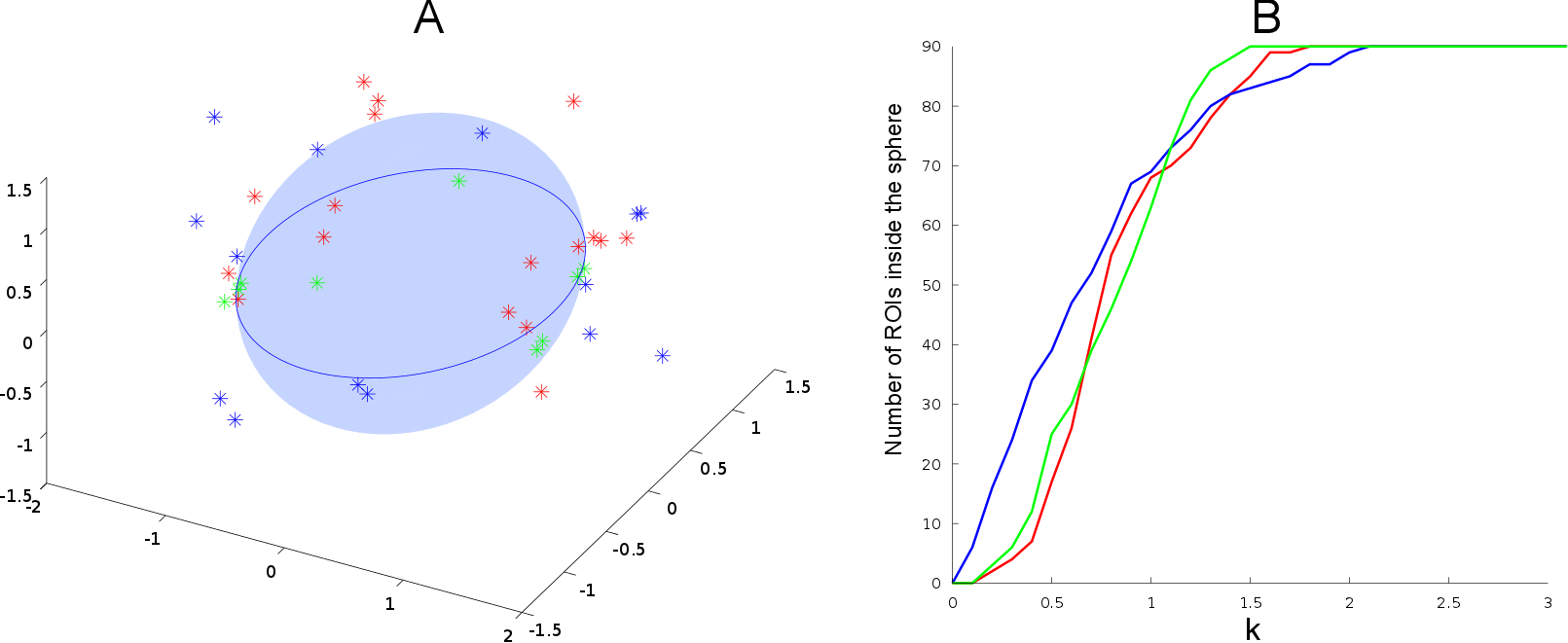}}
\caption{Comparisons between DTI and all other embedded fMRI networks. (A) 3D projections from Eq.~(\ref{hyper-grid}). Only points outside the sphere are plotted. (B) Number of ROI's inside the sphere of radius of $k\times\sigma$.
Results from a single subject, average connectivity and multiplex networks are represented by the red, blue and green points and curves, respectively. We consider the regions statistically different for $k>1$ and statistically equal for $k<1$. This shows that the multiplex algebraic aggregation (green) is more similar algebraically to DTI then average aggregation (blue) and single subject fMRI network (red).  
\label{fig5}}
\end{figure}

Finally, figure~\ref{fig5} displays the difference grid $M$ in a low-dimensional space. As defined in Eq.(~\ref{hyper-grid}),  $M$ corresponds to the relative distance between nodes in networks from different modalities. Differences between brain areas are represented as points widely distributed in the low-dimensional space. Those nodes from different modalities (fMRI and DTI) that share an identical topological structure are located at the origin. The larger the difference in the connectivity structure, the larger the distance from the origin. By setting a threshold $k\times\sigma$, one can identify  brain areas with similar connectivity as those points that lie inside of the hyper-sphere of radius $k\times\sigma$ with centre at the origin.  

The number of brain regions (ROIs) with similar anatomical and functional connectivity are given in figure~\ref{fig5}~(B) as a function of the threshold $k\times\sigma$. Curves correspond to the number of regions inside a hyper-sphere of various radius. We notice that the number of regions differ as a function of the aggregated network's type. It is worthy to mention that the differences above $k$-standard deviations are the important ones, since is above this threshold that the ROI's or nodes become statistical different when compare networks. In our example, the fluctuations below one standard deviations may give us some trend but all nodes in the networks are statistical equal for all types of aggregation. For our specific case, as an example, the brain areas located outside the hyper-sphere of radius $k=1.2$ for the two types of aggregation, are listed in the table 1. 

\begin{table}
\small
  \centering
  \label{tableDifferences}
  \begin{tabular}{@{\vrule height 10.5pt depth3pt  width0pt}l | l | l} \hline
    \multicolumn{3}{ c }{\textbf{Averaged aggregated network}} \\ \hline  \hline
 Calcarine L &  Lingual L  &  Occipital Sup L\\ 
 Calcarine R &  Lingual R &  Occipital Sup R\\ 
 Cuneus L &  Temporal Sup L  &  Occipital Mid L\\ 
 Cuneus R &   Temporal Sup R & Occipital Mid R\\ 
 Occipital Inf &   Insula R &  \\  \hline
    \multicolumn{3}{ c }{\textbf{Multiplex aggregated network}} \\ \hline  \hline
 Cingulum Post R & Occipital Mid R &  Thalamus L \\
 Amygdala R & Occipital Inf L &  Heschl R \\
 Postcentral R & Occipital Inf R &  Temporal Sup R \\
  \end{tabular}
  \caption{ROIs with connectivity differences from DTI at 1.2 standard deviation}
\end{table}

\section{Discussion}
The recent prevalence of applications involving multidimensional and multimodal brain data has increased the demand for technical developments in the analysis of such complex data. Indeed, the discrepancy between structural and functional brain connectivity is a current challenge for understanding general brain functioning.  In this paper, we presented a method for characterising the correspondences between functional and anatomical connectivity. To summarise, the main steps of our method are: 
\begin{enumerate}
\item  Metric network embedding: This procedure embed a group of connectivity graphs in a common space allowing straightforward comparisons. In contrast with simple averaging of connectivity matrices, the  topologically algebraic aggregation can preserve components that are common across different subjects or different neuroimaging modalities. This tensor-based aggregation allows enhancing the common underlying structures providing a natural mathematical framework for studying connectivity data with multidimensional structure. 
\item Multimodal Networks comparison: the differences between the embedded networks are calculated and represented in low-dimensional space. Multi-Dimensional Scaling simply enables to display the information contained in the resulting distance matrix allowing thus an exploratory analysis of the data.
\item Detection of nodes (ROI's) with different connectivities: from points widely distributed in the low-dimensional space one can detect brain nodes that share a similar topological structure as those points are located close to the the origin. One can identify brain areas with the largest difference between anatomical and functional connectivity as those points located outside an imaginary hyper-sphere of a radius given by a threshold (Table 1)
\end{enumerate}


Our findings suggest that embedding a brain network on a metric space may reveal regions that are members of large areas or subsystems rather than regions with a specific role in information processing. This is clearly illustrated for the anatomical network in figure~\ref{fig5} D, where frontal and occipital brain areas of both hemispheres are situated at distantly and located points of the space. Contrary to a classical averaging of connectivity matrices, the embedding of the multiplex functional network reveals brain areas that play a role in large brain system such as the occipital regions, known to be active when the subject is at wakeful rest.

Although experimental evidence suggests that functionally linked brain regions have an underlying structural core, this relationship does not exhibit a simple one-to-one mapping~\cite{wang14}. These correspondences have also been investigated in specific subsystems, must of them focused on the default mode network (DMN), which is a group of brain regions that preferentially activate when individuals engage in internal tasks, i.e. when the subject is not focused on the outside world but the brain is at wakeful rest. Several studies report that the DMN exhibits a high overlap in its structural and functional connectivity~\cite{wang14, honney09}. Nevertheless, strong discrepancies have been reported and strong functional links can be found between regions without direct structural linkages~\cite{honney09}. 
 
At a group level, one of the reasons for this discrepancy between structural and functional connectivity has been suggested to be the functional variability across subjects~\cite{skudlarski08, honney09, wang14}. Indeed, clinical studies have provided evidence for a large heterogeneity of the functional connectivity, particularly in groups of patients with brain disorders such as neuropsychiatric disorders, which strongly alters the structural-functional relationships~\cite{wang14}. Analytical tools are therefore required to account for this variability in order to enhance the common underlying network structure.

Results suggest that averaged aggregation captures the general differences in regions that play a role in visual, auditory and body self-awareness processes, but fails to identify in detail other specific areas \emph{across} the subjects/groups. In table 1 we observed that the average aggregation essentially captures part of visual (calcarine,  cuneus, lingual, occipital), auditory  (superior temporal gyrus), and insula regions that are associated to visual process and body self-awareness. We  expected in average that these regions to be highly activated during the scans. This is alined with the fact that rs-fMRI was acquired with closed eyes and the subjects have some auditory and body awareness. 

From the multiplex aggregation (or algebraic aggregation) shown in table 1, we observed that besides capturing the well known visual (occipital areas), primary sensory cortex (postcentral) and auditory regions (Heschl gyrus, superior temporal, thalamus), this approach also captures some other network sub-structures involved in touch activation (postcentral gyrus, thalamus) and emotional state activations (amygdala, thalamus,  posterior cingulate). This alines with our claim that algebraic aggregation preserves better the multilayer sub-structures across a group of subjects (multilayers) accounting for as much of the variability in the data as possible. 

Although we cannot definitively a one-to-one mapping of the structural and functional connectivity, we think that our method could provide new insights on the organisation of brain networks during diverse cognitive or pathological states. We therefore hope that our approach will foster more principled and successful analysis of multimodal brain connectivity datasets.

For all the methods described in this article we provide the corresponding MATLAB software code. Data and code are freely available at  the website \url{https://sites.google.com/site/fr2eborn/download}

\section*{Acknowledgements}
The authors thank Y. Iturria-Medina and L. Melie-Garc\'{\i}a for sharing the DTI connectivity data used in the study. This work was supported by the LASAGNE (Contract No.318132) and MULTIPLEX (Contract No.317532) EU projects. A.D.-G. acknowledges support from Generalitat de Catalunya (2014SGR608) and Spanish MINECO (FIS2012-38266).


\end{document}